\documentclass[conference]{IEEEtran}
\IEEEoverridecommandlockouts
\usepackage{svg}
\usepackage{soul}
\usepackage{cite}
\usepackage{amsmath,amssymb,amsfonts}
\usepackage{graphicx}
\usepackage{textcomp}
\usepackage{xcolor}
\usepackage{caption}
\usepackage{subcaption}
\usepackage[left=0.680in, right=0.620in, bottom = 1in, top=0.7in]{geometry}
\usepackage{url,amssymb,threeparttable,multirow,booktabs, tabularx}
\def\BibTeX{{\rm B\kern-.05em{\sc i\kern-.025em b}\kern -.08em
    T\kern-.1667em\lower.7ex\hbox{E}\kern-.125emX}}
\usepackage{algpseudocode} 
\usepackage{algorithm}

\begin{document}

\title{Scaling 6G Subscribers with Fewer BS Antennas using Multi-carrier NOMA in Fixed Wireless Access\\
}

\author{
\IEEEauthorblockN{Kamyar Rajabalifardi\textsuperscript{1}, Sagnik Bhattacharya\textsuperscript{1}, Mehrnaz Afshang\textsuperscript{2}, Mohammad Mozaffari\textsuperscript{2}, John M. Cioffi\textsuperscript{1}}

\IEEEauthorblockA{\textsuperscript{1}Dept. of Electrical Engineering, Stanford University, Stanford, CA, USA} 

\IEEEauthorblockA{\textsuperscript{2}Ericsson Research, Santa Clara, CA, USA} 

\IEEEauthorblockA{
Emails: \{kfardi, sagnikb, cioffi\}@stanford.edu, \{mehrnaz.afshang, mohammad.mozaffari\}@ericsson.com}
}

\maketitle

\begin{abstract}
This paper introduces a novel power allocation and subcarrier optimization algorithm tailored for fixed wireless access (FWA) networks operating under low-rank channel conditions, where the number of subscriber antennas far exceeds those at the base station (BS). As FWA networks grow to support more users, traditional approaches like orthogonal multiple access (OMA) and non-orthogonal multiple access (NOMA) struggle to maintain high data rates and energy efficiency due to the limited degrees of freedom in low-rank scenarios. Our proposed solution addresses this by combining optimal power-subcarrier allocation with an adaptive time-sharing algorithm that dynamically adjusts decoding orders to optimize performance across multiple users. The algorithm leverages a generalized decision feedback equalizer (GDFE) approach to effectively manage inter-symbol interference and crosstalk, leading to superior data rates and energy savings. Simulation results demonstrate that our approach significantly outperforms existing OMA and NOMA baselines, particularly in low-rank conditions, with substantial gains in both data rate and energy efficiency. The findings highlight the potential of this method to meet the growing demand for scalable, high-performance FWA networks.
\end{abstract}


\section{Introduction}
\label{sec:intro}
The recent exponential growth in fixed wireless internet access (FWA) subscription exceeds all projections. FWA offers high-speed internet access from an installed fixed wireless base station (BS). FWA has emerged as the primary efficient and cost-effective alternative to laying optical fiber cables, especially in locations with challenging topology. FWA is projected to be the primary enabler of the high data rate and low power consumption requirements as users shift to applications like mixed reality (MR), distributed machine learning, federated learning, autonomous vehicle networks, etc. in 6G communications. According to industry forecasts \cite{Ericsson2023FWA}, FWA subscriptions are expected to soar, with millions of users worldwide embracing this technology for its reliability, speed, and the flexibility it offers over traditional broadband methods. However, as thousands of users subscribe to an FWA base station (BS), we hit a fundamental bottleneck; the disparity between the number of antennas at the base station (BS) and the exponentially increasing number of subscriber antennas. We are faced with an extreme low-rank channel scenario, with the number of participating users at any given time is far greater than the number of AP antennas. In this situation, the degrees of freedom available in the channel for transmitting signals is significantly less than the demand posed by the multitude of subscriber antennas. This emerging bottleneck is a critical concern for the scalability and efficiency of FWA networks, especially with the high data rate requirements.

Traditional solutions using orthogonal multiple access of frequency-time-space resources (OMA), like Multiple-Input Multiple-Output (MIMO) and massive MIMO systems \cite{chang1966synthesis}, which thrive on increasing the number of antennas at the base station (BS) to boost capacity, are impractical for this scenario. These methods inherently rely on the number of transmit antennas being equal to or more than the number of receive antennas, which is impractical with thousands of FWA users connected to the same BS. The low-rank nature of the channel in such a setup makes conventional OMA techniques, like those utilized in current 5G or Wi-Fi systems, result in low data rates while consuming high energy.

It has been demonstrated that power allocation techniques used in non-orthogonal multiple access (NOMA) result in higher data rates and reduced power consumption compared to orthogonal multiple access (OMA) methods \cite{ tse_viswanath_2005}. The study in \cite{noma6} examines NOMA for single-input-single-output (SISO) systems, where the optimal order for successive interference cancellation (SIC) is determined simply by the channel strength (or channel coefficient) between each user and BS. However, in contrast to SISO, multi-user MIMO scenarios do not naturally provide an order based on users’ channels \cite{noma7}. This lack of inherent ordering necessitates various heuristic approaches for determining the decoding order in MIMO scenarios, such as assuming only two users in the system \cite{noma8}, grouping users into pairs and applying NOMA within each group \cite{noma9}, among others. These heuristic methods result in data rates and power consumption levels that do not meet current performance requirements. Furthermore, the aforementioned studies assume that the entire bandwidth is allocated to all users, with multiple access achieved solely through power or code domains, contributing to less-than-optimal performance. In \cite{chai2023iterative}, the authors investigate multi-carrier hybrid NOMA (MC-NOMA), where both power and subcarrier allocation are optimized for downlink transmissions. However, they also rely on a heuristic decoding order at each user, determined by the absolute value of the channel coefficient for that user. This approach is not valid when the channel is represented as a vector for systems with multiple antennas at the BS and/or at each user.

Addressing the shortcomings of the current work, this paper introduces a novel optimal power-subcarrier allocation for data rate maximization and power consumption minimization for FWA networks. The proposed algorithm especially outperforms the OMA and NOMA methods by a significant margin in low rank channel scenarios. For the first time, we tackle the scenario where the number of antennas at the BS is vastly outnumbered by the antennas at subscriber homes, thus presenting an extremely low rank channel. The proposed algorithm derives the optimal decoding order for SIC, instead of making heuristic assumptions on the decoding order. The proposed algorithm further develops and implements a novel time-sharing algorithm, which uses adaptive decoding order varying to achieve higher data rates than any single decoding order can. The proposed method not only confronts the low-rank channel bottleneck head-on but also achieves optimal data rates and energy efficiency under those constraints, which is crucial for emerging applications demanding substantial bandwidth, such as AR/VR, distributed generative modeling, and rendering. This novel contribution involving GDFE-based data rate and energy optimization for low rank FWA scenarios sets the stage for enabling a more robust, scalable FWA network. Our extensive simulations show that the proposed algorithm achieves higher sum datarate, especially under low rank channel conditions, compared to OMA and NOMA baselines respectively.

\begin{figure}[t]
    \centering
    \includegraphics[trim = {100, 45, 80, 75}, clip, height = 4.8cm]{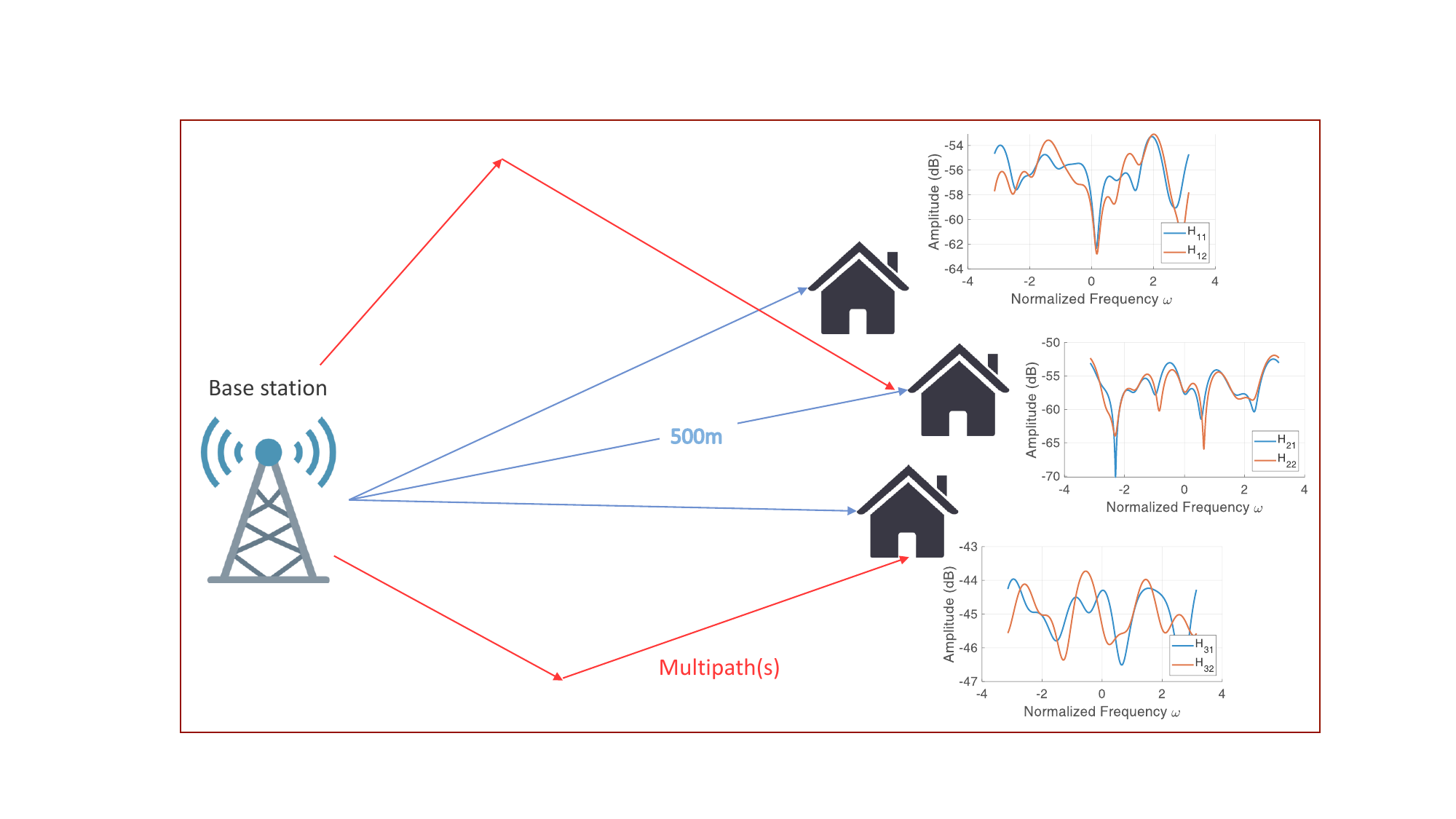}
    \caption{Fixed Wireless Access Scenario}
    \label{fig:fwabc}
\end{figure}
\section{Methodology}
\label{sec:methodology}

Fig.~\ref{fig:fwabc}
depicts the downlink broadcast channel (BC), in which a base station (BS) equipped with $n_T$ antennas transmits signals to a group of $U$ users. Each $u^{th}$ user is equipped with $n_{R, u}$ antennas and receives a symbol vector $\boldsymbol{\bar{x}}_u \in \mathbb{R}^{n_{T,u}}$ from the BS. The received signal at the user's end, denoted by $\boldsymbol{y}$, can be expressed as follows:

\begin{equation} \label{eq:MIMO}
    \boldsymbol{\bar{y}} = H_{BC} \:.\: \boldsymbol{\bar{x}} + \boldsymbol{\bar{n}}
\end{equation}
where $\boldsymbol{\bar{x}} \in \mathbb{R}^{n_T}$ represent the concatenation of all symbols $\boldsymbol{\bar{x}}_u$ transmitted by the BS to the receivers, and $\boldsymbol{y} \in \mathbb{C}^{n_R}$ be the aggregated received signal across all users, where $n_R = \sum_{u=1}^{U} n_{R,u}$. The channel matrix is denoted by $H_{BC} \in \mathbb{C}^{n_R \times n_T}$, which captures the transmission dynamics from the BS to each user. Additionally, $\boldsymbol{\bar{n}} \in \mathbb{R}^{n_R}$ represents the additive white Gaussian noise (AWGN) affecting the reception (we will use notations $\boldsymbol{x}, \boldsymbol{n}, \boldsymbol{y}$ for MAC direction in subsection \ref{subsec:duality}). In the context of a multi-carrier system with $N$ sub-carriers, Equation~\eqref{eq:MIMO}
 can be separately formulated for each sub-carrier.

In Orthogonal Multiple Access (OMA) \cite{6495771}, \cite{6762926} approach different users are allocated distinct non-overlapping frequency or time resources to prevent interference among them. Traditional linear receivers \cite{chang1966synthesis} utilize a technique known as simultaneous water-filling (SWF) \cite{book} to maximize the sum of data rates within the available energy constraints. However, this approach often results in suboptimal data rates and energy efficiency, especially in scenarios where leveraging crosstalk between users could instead enhance data rates, rather than merely treating it as noise to be minimized for each user.

In downlink NOMA approach \cite{islam2016power}, a BS transmits a combination of users' messages using superposition coding (SC), and users at the receiver side decode their messages using the successive interference cancellation (SIC), which requires a decoding order implying which user should be decoded first. 

If $\boldsymbol{\pi}(.)$ represents the decoding order vector, such that $\boldsymbol{\pi}(1)$ is the decoding order for user 1, and so on, then, $\boldsymbol{\pi}^{-1}(.)$ is defined as the inverse of $\boldsymbol{\pi}(.)$, where $\boldsymbol{\pi}^{-1}(1)$ indicates the user index decoded first, the achieved data rate of user $\boldsymbol{\pi}^{-1}(u)$ can be expressed as \cite{book}:
\begin{equation}\label{eq:SIC}
    \begin{aligned}
    &b_{\boldsymbol{\pi}^{-1}(u)} = \sum_{i=1}^u b_{\boldsymbol{\pi}^{-1}(i)} - \sum_{i=1}^{u-1} b_{\boldsymbol{\pi}^{-1}(i)} \\
    &\footnotesize{= \operatorname{log}_2 \left|\frac{R_{\boldsymbol{\bar{n}\bar{n}}} + \sum_{i=1}^u H_{BC}({\boldsymbol{\pi}^{-1}(i)})\cdot R_{\boldsymbol{\bar{x}\bar{x}}}(\boldsymbol{\pi}^{-1}(i))\cdot H_{BC}^*({\boldsymbol{\pi}^{-1}(i)})}{R_{\boldsymbol{\bar{n}\bar{n}}} + \sum_{i=1}^{u-1} H_{BC}({\boldsymbol{\pi}^{-1}(i)})\cdot R_{\boldsymbol{\bar{x}\bar{x}}}(\boldsymbol{\pi}^{-1}(i))\cdot H_{BC}^*({\boldsymbol{\pi}^{-1}(i)})} \right|}
    \end{aligned}
\end{equation}
where $R_{\boldsymbol{\bar{n}\bar{n}}} \in \mathbb{R}^{n_R\times n_R}$ is the noise auto-correlation matrix, and $H_{BC}({\boldsymbol{\pi}^{-1}(u)})$ represents the channel between BS and user $\boldsymbol{\pi}^{-1}(u)$. 

In NOMA, channel state information (CSI) determines the deconding order $\mathbf{\pi}$\cite{islam2016power}. In particular, if the users' channel gain are sorted as:
\begin{equation}
    0 < \left\|H_1 \right\|^2 \leq \left\|H_2 \right\|^2 \leq \cdots \leq \left\|H_U \right\|^2
\end{equation}
then user 1 is decoded first, and its signal is subtracted from the composite signal to facilitate the decoding of user 2's message, and this process continues sequentially. Although this method has proven effective in numerous scenarios, relying solely on CSI to determine the decoding order may lead to a suboptimal solution in some cases.

In following we present an optimal power allocation scheme for various subcarriers in a multiple access channel (MAC) scenario, where users transmit data to a base station (BS). We leverage the duality between broadcast channels (BC) and MAC problems. Specifically, if we can solve the power allocation in one scenario, we can address a similar problem in the other scenario, albeit with different parameters such as autocorrelation and channel matrices, as established by  \cite{jindal2004duality}.

\subsection{MAC2BC and BC2MAC} \label{subsec:duality}
The mathematical formulation explained in subsection \ref{Esum} is dedicated to the multiple access channel (MAC). However, the problem we are solving pertains to the broadcast channel (BC). It can be proven that there is a duality between BC and MAC scenarios, allowing problems in one scenario to be transformed interchangeably into the other \cite{jindal2004duality}. In particular, if $H_{\textrm{BC}} \in \mathbb{R}^{n_R \times n_T}$ is the channel matrix of BC, the channel matrix of MAC $H_{\textrm{MAC}}\in \mathbb{R}^{n_T \times n_R}$ can be expressed as \cite{book}:
\begin{equation} \label{eq:hbc}
    H_{\text {MAC}}=\mathcal{P}_T \cdot H_{BC}^* \cdot \mathcal{P}_R
\end{equation}
where $\mathcal{P}_T \in \mathbb{R}^{n_T \times n_T}$ and $\mathcal{P}_R \in \mathbb{R}^{n_R\times n_R}$ are permutation matrices defined as follows:
\begin{equation}
\mathcal{P}_{T} \triangleq\left[\begin{array}{ccc}
0 & 0 & I_{n_{T, 1}} \\
0 & . \cdot & 0 \\
I_{n_{T, U}} & 0 & 0
\end{array}\right], \: \mathcal{P}_{R}\triangleq\left[\begin{array}{ccc}
0 & 0 & I_{n_{R, U}} \\
0 & . \cdot & 0 \\
I_{n_{R, 1}} & 0 & 0
\end{array}\right]
\end{equation}
where $I_{n_{T,u}}$ and $I_{n_{R,u}}$ are identity matrices.

\begin{figure*}[t!]
    \centering
    \includegraphics[trim = {10, 5, 4, 10}, clip, height = 1.65cm]{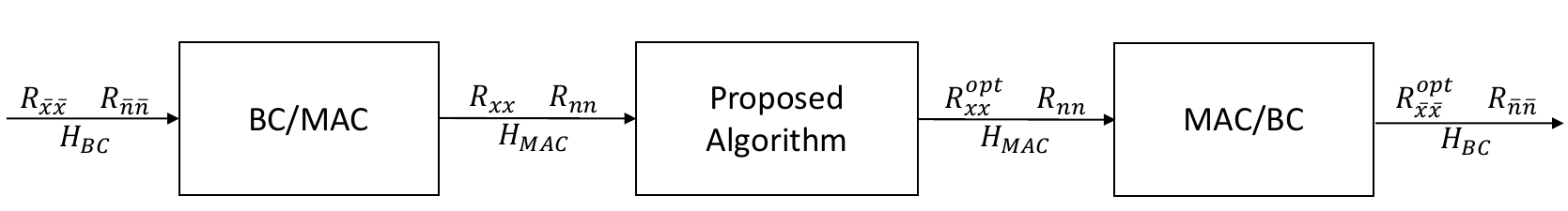}
    \caption{Block Diagram of the duality between MAC and BC}
    \label{fig:blockdiag}
\end{figure*}

Fig.  \ref{fig:blockdiag} illustrates depicts the whole process of power allocation in which $\boldsymbol{x}$ and $\boldsymbol{n}$ are the symbol vector and noise vector used in MAC and $\bar{\boldsymbol{x}}$ and $\bar{\boldsymbol{n}}$ are the symbols and noise vector used in BC design respectively. Chapter 5, section 5.5.1.1 of \cite{book} explains the MAC2BC and BC2MAC algorithms and the transformation of $R_{\boldsymbol{x}\boldsymbol{x}}$ to $R_{\boldsymbol{\bar{x}}\boldsymbol{\bar{x}}}$ and vice versa in details.

\subsection{Optimal Power Allocation Problem} \label{Esum}
The weighted energy-sum minimization is formulated as follows:
\begin{equation}
\label{eq:energy-sum-minimize}
\begin{aligned}
&\min_{\left\{R_{\boldsymbol{xx}}{(u, n)},\:b_{u,n} \right\}} \quad \sum_{u=1}^U \sum_{n=1}^N w_u \cdot \operatorname{trace}\left\{R_{\boldsymbol{x x}}(u, n)\right\}\\
&\textrm{subject to} \quad\\ &\mathbf{b}=\sum_{n=1}^N\left[b_{1, n}, b_{2, n}, \ldots, b_{U, n}\right]^T \succeq \mathbf{b}_{\min } \succeq \mathbf{0}\\ \\
&\small{\sum_{u \in T} \sum_{n=1}^N b_{u, n} \leq \log _2\left|\frac{R_{n \boldsymbol{n}}+\left(\sum_{u \in T} H_{u, n} \cdot R_{x x}(u, n) \cdot H_{u, n}^*\right)}{R_{n \boldsymbol{n}}}\right|}\\
&\qquad \qquad \forall \: T  \subseteq \{1, 2, \cdots, U\} \\
\end{aligned}
\end{equation}
\vspace{0.02in}
where $R_{\boldsymbol{xx}}(u)$ and $b_{u,\operatorname{min}}$ are the autocorrelation matrix and user $u$'s minimum data rate, respectively. Also, the vector $w \in \mathbb{R}_{0+}^U$ represents non-negative weights for each user's energy. The first constraint forces the final data rates to not be less than the minimum desired data rate for each user. Moreover, the second set of constraints are based on the Shannon's capacity formula for multi-user case \cite{shannon}. In other words, for any subset of users, denoted as $\{1, 2, \cdots, U\}$, the overall data rate does not exceed the channel capacity. Since the objective function is convex and all constraints represent convex sets, this problem can be solved using CVX package \cite{cvx}. 

It can be proven that the optimal decoding order can be derived from the optimization problem (\ref{eq:energy-sum-minimize}) \cite{book}. In particular, it is shown that the lagrange multipliers corresponding to the minimum required data rates constraints (second set of constraints in problem (\ref{eq:energy-sum-minimize})) represents the optimal decoding order. If $\theta_u$ (for all $u \in \{1, 2, \cdots, U\}$) is the lagrange multiplier corresponding to the constraint $\sum_{n=1}^N b_{u,n} \geq \mathbf{b}_{min, u}$, then the optimal decoding order $\pi$ must satisfy the inequality below:
\begin{equation}
\label{eq:decoding_order}
    \theta_{\mathbf{\pi}^{-1}(U)} \geq \theta_{\mathbf{\pi}^{-1}(U-1)} \geq ... \geq \theta_{\mathbf{\pi}^{-1}(1)}
\end{equation}
Section 5.4 of \cite{book} provides the detailed proof of the aforementioned inequality. Therefore, the decoding order can be derived from the lagrange multipliers corresponding to the minimum required data rate constraints in the energy minimization problem.

\subsection{Time Sharing}
\label{subsec:time-share}
The proposed time-sharing algorithm tackles scenarios with identical $\theta_i$ values (Lagrangian multipliers for data rate constraints) among users, by first identifying users with matching $\theta$ values and grouping them into clusters. This crucial step helps determine candidates for time-sharing based on their data rate goals and potential for similar resource sharing. For each cluster, permutations of user sequences are considered to explore different resource distribution possibilities and achieve target rates. The algorithm reviews each permutation for its achievable rates and energy use in a specified decoding order. After assessing all permutations, it constructs the convex hull of the $\theta$ configurations to pinpoint optimal boundary points. Given the objective function's reliance on linear combinations of $\theta_i$, focusing on linear combinations of the convex hull boundary points suffices. The algorithm finally deduces the time fractions for each symbol period necessary to achieve target rates by solving a linear system, with these fractions as variables and the target rates as objectives.

\section{Baseline Methods}
\label{sec:baselines}

We compare the performance of the proposed algorithm for optimal power-subcarrier allocation and time sharing with three baseline methods widely used in related work. The first baseline is OMA method used in current Wi-Fi standards \cite{6762926}, which allocated disjoint frequency time resource blocks to participating users. OMA uses a linear receiver, where every user signal is decoded considering all other interfering signals as noise. The second baseline is NOMA power allocation \cite{noma1}, which allocates the whole bandwidth to all participating users. NOMA does an optimal power allocation by consdering a fixed decoding order based on channel strength ordering for all users. Lastly, the final baseline used is multi-carrier NOMA (MC-NOMA) power-subcarrier allocation \cite{mc-noma}. MC-NOMA does an optimal power allocation for each subcarrier, similar to the proposed algorithm. However, it makes a heuristic assumption for the decoding order based on the channel strength order for the participating users. 
\section{Experiments}
\label{sec:experiments}
In this section, we experiment our method and compare it with other methods. such as OMA. For our simulations, we use QuaDRiGa \cite{jaeckel2014quadriga}, a MATLAB software tool used for channel modeling. 
\subsection{FWA Channel Model} \label{subsec:channelmodel}
\begin{figure}[h]
    \centering
    \includegraphics[trim = {20, 180, 20, 210}, clip, height = 5.7cm]{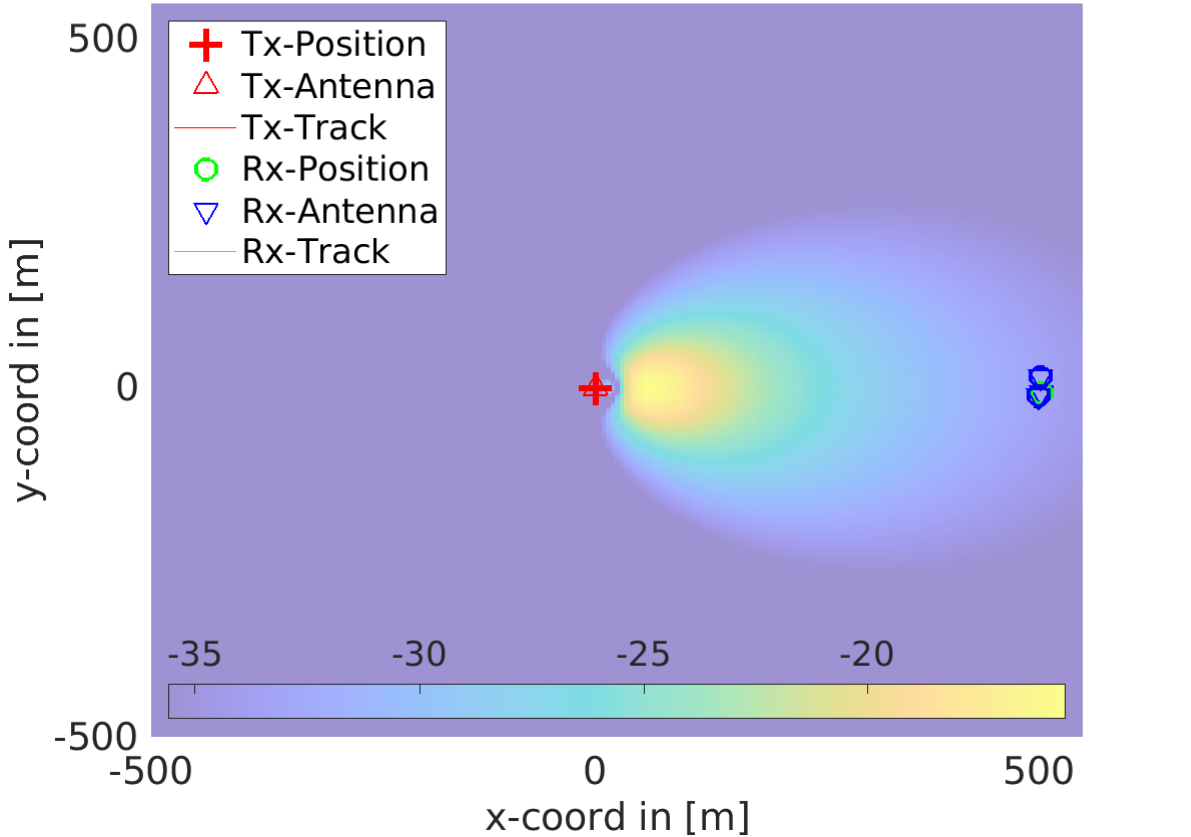}
    \caption{Received power [dBm] for 3.5GHz, and for transmission power of 50dBm}
    \label{fig:power-map}
\end{figure}

For channel modeling, we utilize the 3GPP-3D antenna configuration \cite{mondal20153d} for BS, a standardized method for simulating the radiation patterns of antennas in wireless communication systems. Unlike traditional 2D models that only consider azimuth (horizontal) patterns, the 3D model incorporates both azimuth and elevation (vertical) patterns. Key parameters include antenna gain, polarization, electrical and mechanical downtilts, and azimuth and elevation beamwidths. These parameters can be specified using variables $A_{in}$, $B_{in}$, $C_{in}$, $D_{in}$, $E_{in}$, $F_{in}$ in QuaDRiGa. This approach offers a more accurate representation of real-world antenna radiation characteristics. Also, for receiver side, we assume each user is equipped with a single ($n_{R, u}=1$) omni-directional antenna. Additionally, our channel modeling framework incorporates the rural macro cell (RMa) scenario from the 3GPP TR 38.901 \cite{docomo20165g}, which provides empirical data and models for simulating wireless network behavior in rural environments, further enhancing the realism and applicability of our simulations. The settings used in our simulations are summarized in Table \ref{tab:experiment-parameters}. Fig. \ref{fig:power-map} illustrates the power map of the 3GPP-3D antenna model with an average distance of 500m between users and the BS antenna.

\begin{table}[t]
    \caption{Experiment Parameters}
    \centering
    \renewcommand{\arraystretch}{1.35}
    \resizebox{7cm}{!}{
    \begin{tabular}{|c|c|c|}
    \hline
      \textbf{Notation} & \textbf{Description} & \textbf{Value} \\
    \hline
      $U$ & Number of users & 3 \\
      $n_{R, u}$ & Number of antennas per user & 1 \\
      $n_T$ & Number of antennas at BS  & 2 \\
      $h_{x}$ & BS height & 30m \\
      $h_{u}$ & UE height & 6m \\
      $d$ & Average distance of UE to BS & 500m \\
      $W$ & Channel bandwidth  & 100MHz \\
      $f_c$ & Carrier frequency  & 3.5GHz \\
      $\lambda$ & Wavelength & $\frac{1}{f_c}$ \\
      $N$ & Number of subcarriers  & 64 \\
      $P_T$ & Transmission power & 50dBm \\
      $\sigma_n^2$ & Noise power spectral density & -174dBm \\
      $W \sigma_n^2$ & Noise power & -94dBm \\
      $G_x$ & BS antenna gain & 13dBi \\
      $G_y$ & user's antenna gain & 0dBi\\
      $A_{in}$ & Number of vertical elements  & 4 \\
      $B_{in}$ & Number of horizontal elements  & 2 \\
      $D_{in}$ & Polarization indicator  & 4 \\
      $E_{in}$ & Electric downtilt angle [deg]  & 12 \\
      $F_{in}$ & Element spacing  & $0.5\lambda$ \\
    \hline
    \end{tabular}}
    \label{tab:experiment-parameters}
\end{table}

\subsection{The effect of changing SNR}
\begin{figure}[t]
    \centering
    \includegraphics[trim = {0, 180, 0, 180}, clip, height = 5.5cm]{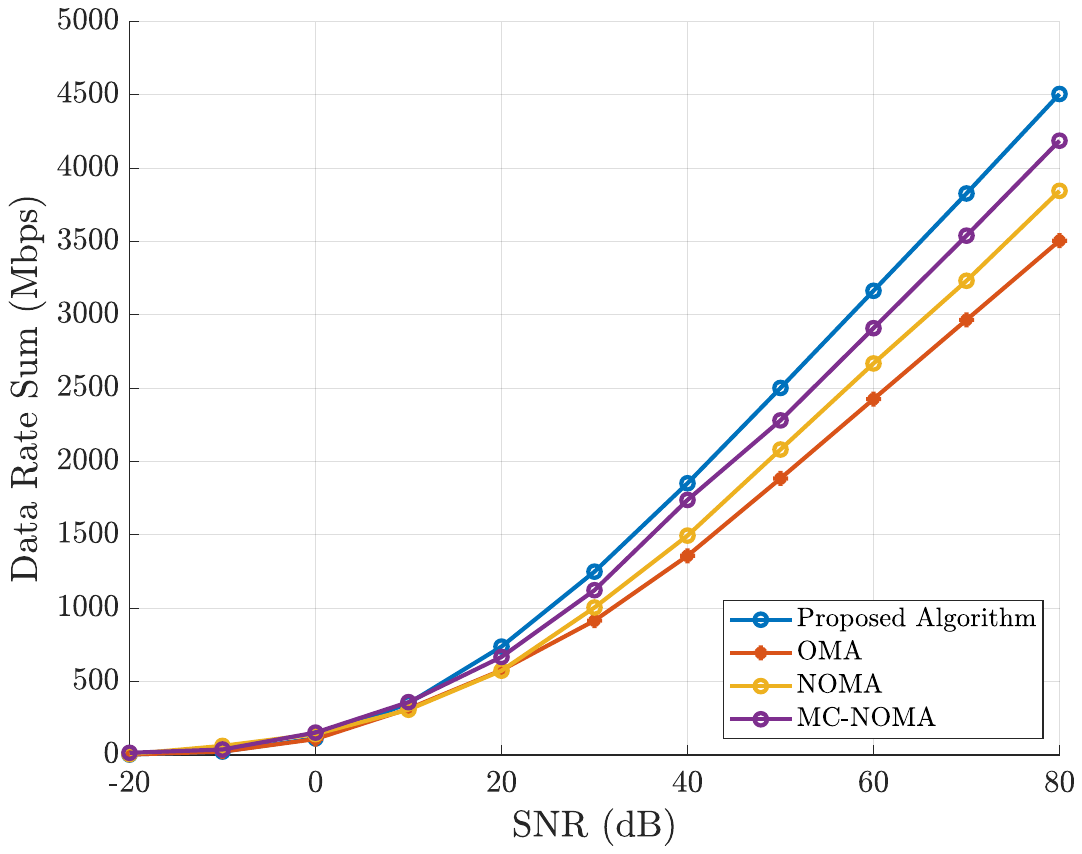}
    \caption{Data rate sum vs receive SNR for 3 users with distance of 500m from BS}
    \label{fig:EvsB}
\end{figure}

Fig. \ref{fig:EvsB} shows the data rate sum as a function of receive SNR for 3 users with 500m distance from BS. As we can see, OMA which utilizes linear receiver, has the worst performance compared to other methods since it does not use the crosstalk among users. For lower SNRs, all methods perform almost the same. This is due to the fact that for low SNRs, the crosstalk effect is not significant. For higher SNRs, the proposed algorithm outperforms NOMA and MC-NOMA since these two baselines consider the CSI as the decoding order, while our proposed algorithm finds the optimal decoding order.

\subsection{The effect of changing distance from BS}
Table~\ref{tab:energy-compare2} shows the transmit power values required by the proposed algorithm to achieve the same data rates as achieved by the OMA-based linear receiver, while increasing users' distance from BS. For this comparison, the OMA power is held constant at 20 dB per user. Initially, the data rates for the linear receiver under these power constraints are computed. Subsequently, the power levels required by the proposed algorithm to match these data rates are calculated. At a distance of 500 meters, the proposed algorithm demonstrates a significant reduction in power consumption, averaging 54\% lower than that of the linear receiver. This efficiency gain stems from the proposed algorithm’s ability to effectively leverage signal crosstalk among different users. This is done by optimizing both the decoding sequence and time sharing strategies, thereby achieving higher data rates than OMA.

\begin{table}[t!]
\caption{Transmit power consumption versus the distance of users from BS for Single Antenna Per User, Fixed energy for linear receiver, $U = 3$, $n_{R, u} = 1$,  $n_{T}=2$, Transmit Power values are relative to 20 dB}
\centering
\resizebox{9.1cm}{!}{
\begin{tabular}{|c|c|c|c|c|}
\hline
\multirow{2}{*}{\begin{tabular}[c]{@{}c@{}}\textbf{Distance} \\ \textbf{from BS (km)}\end{tabular}} & \multicolumn{2}{c|}{\textbf{OMA}} & \multicolumn{2}{c|}{\textbf{Proposed Algorithm}} \\ \cline{2-5} 
                     & \textbf{Transmit Power}& \textbf{Data Rates (Mbps)}& \textbf{Transmit Power}& \textbf{Data Rates (Mbps)  }\\ \hline
0.5                    & [1,  1,  1] & [114, 204, 138]& [0.46, 0.46, 0.46]& [114, 204, 138]\\ \hline
1                    & [1,  1,  1] & [91, 25, 105]& [0.79, 0.77, 0.78]& [91, 25, 105]\\ \hline
1.5                    & [1,  1,  1] & [8, 53, 27]& [0.86, 0.80, 0.82]& [8, 53, 27]\\ \hline
2                    & [1,  1,  1] & [17, 18, 2]& [0.88, 0.92, 0.95]& [17, 18, 2]\\ \hline
\end{tabular}}
\label{tab:energy-compare2}
\end{table}

\subsection{The effect of changing $n_T$}
\begin{figure}[t]
    \centering
    \includegraphics[trim = {0, 180, 0, 180}, clip, height = 5.5cm]{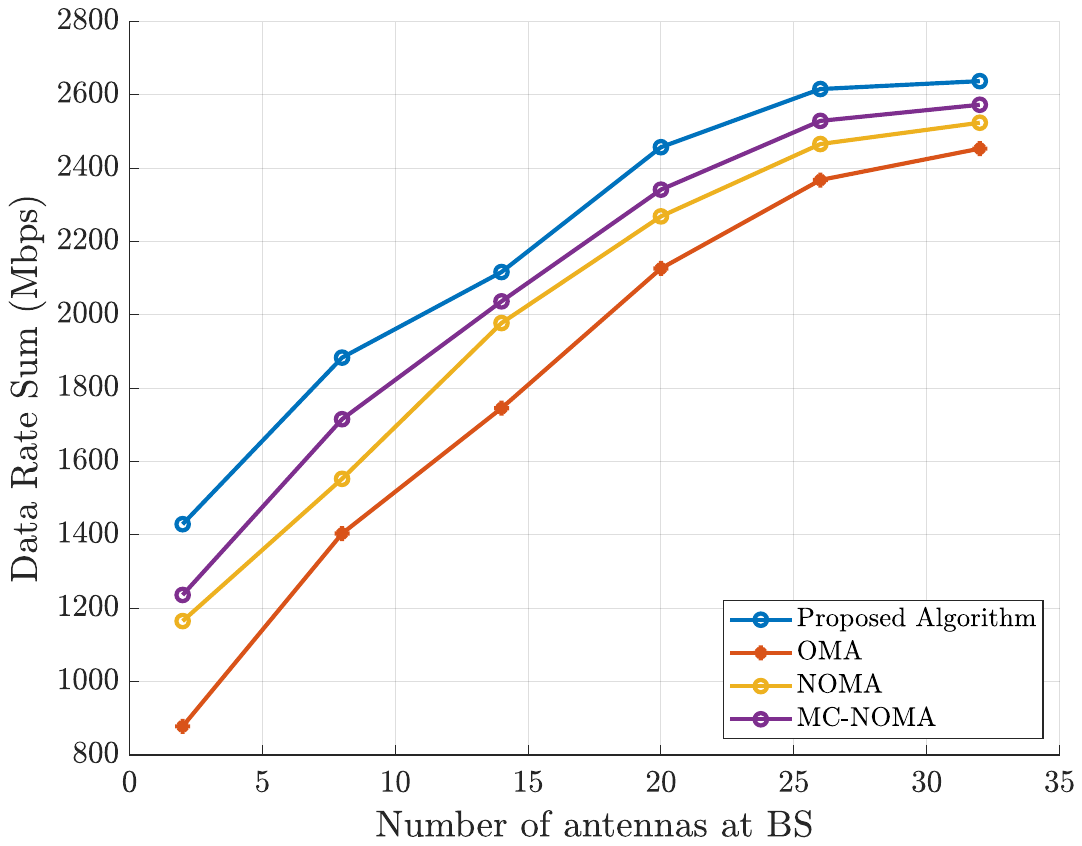}
    \caption{Data rate sum vs Number of antennas at BS for 3 users with average distance of 500m from BS}
    \label{fig:changeAP}
\end{figure}

It can be demonstrated that as the number of antennas at the base station (BS) increases, the precoder's triangular signal processing matrix, which resembles successive interference cancellation (SIC), becomes diagonal. This suggests that feedback at the transmitter is not advantageous, and orthogonal multiple access (OMA) provides performances almost as good as feedback equalization \cite{book}. Fig.  \ref{fig:changeAP} illustrates the effect of increasing $n_T$ while keeping $n_{R, u} = 1$. As shown, with an increase in the number of antennas at the BS, the data rate sums achieved using the proposed algorithm, NOMA, MC-NOMA and OMA converge, indicating that Non-orthogonal methods and OMA perform comparably in massive MIMO scenarios. We do want to note that the classic FWA scenario, with thousands of users communicating with a single AP with less antennas, will be a very low rank channel, and thus this massive MIMO advantage will not be there for those cases. In that case, the proposed algorithm provides significantly higher data rates compared to OMA.

\subsection{The effect of crosstalk}\label{subsec:crosstalk}
\begin{figure}[t]
    \centering
    \includegraphics[trim = {0, 180, 0, 180}, clip, height = 5.5cm]{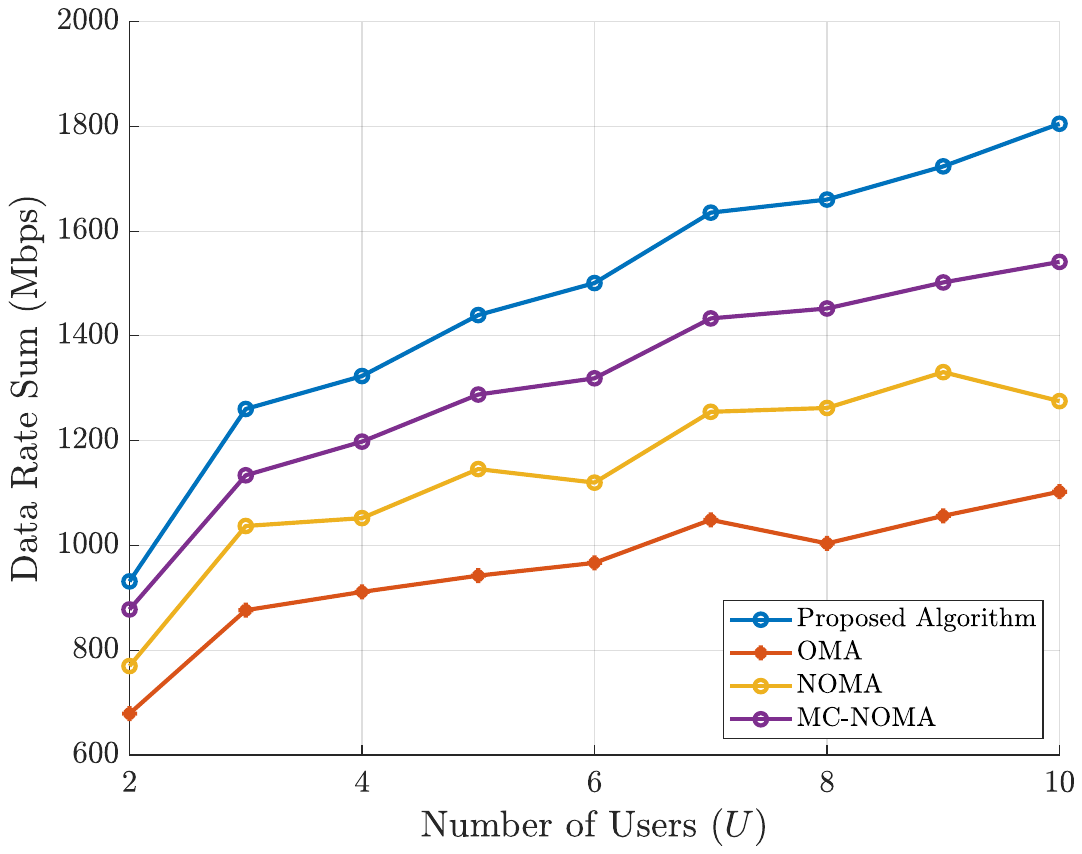}
    \caption{Data rate sum vs Number of Users with distance of 500m from BS}
    \label{fig:crosstalk}
\end{figure}

Fig. \ref{fig:crosstalk} illustrates the impact of crosstalk among different numbers of users. In this experiment, users have an average distance of 500m from BS, and are placed in close proximity to each other to enhance the crosstalk effect. As shown, increasing the number of users widens the gap between the data rate sum derived from Orthogonal Multiple Access (OMA) and that from the proposed algorithm. This demonstrates that the proposed approach effectively makes use of the cross talk among users more effectively.

\subsection{The effect of time sharing}
\begin{figure}[t]
     \centering
     \hspace{-0.8cm}
     \begin{subfigure}{0.2\textwidth}
         \centering
         \includegraphics[trim = {45, 185, 65, 205}, clip, height = 3.6cm]{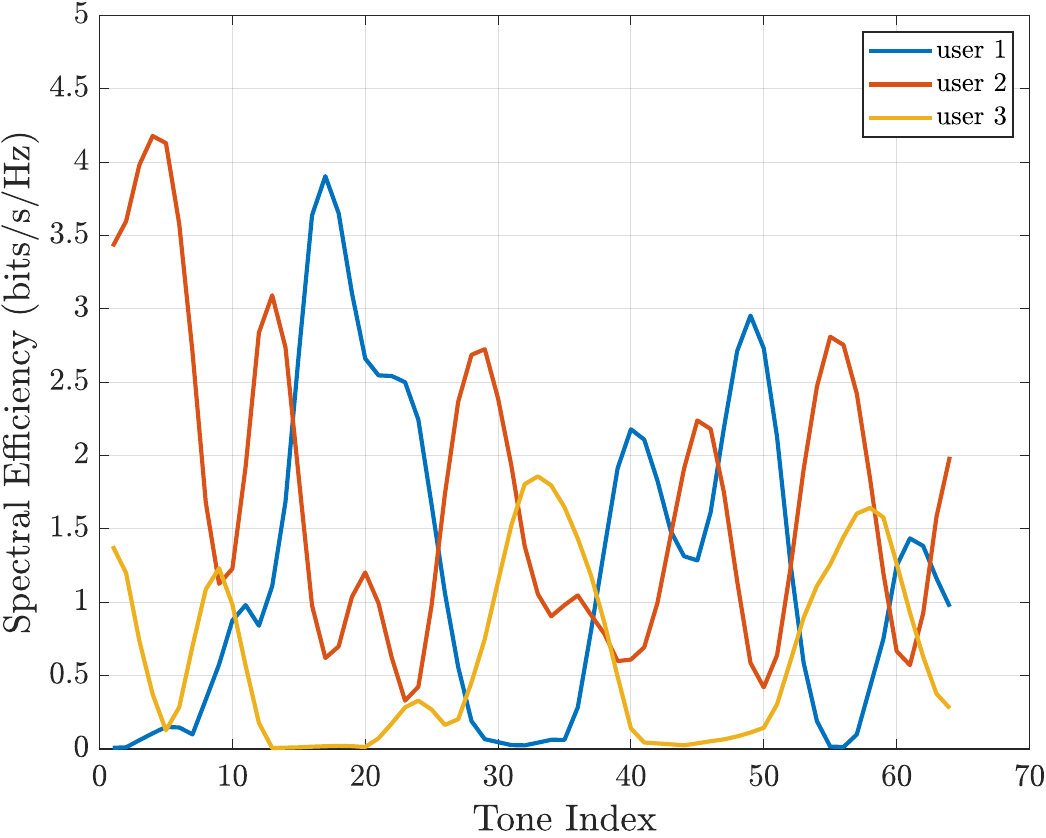}
         \caption{$1^{st}$ time slot (91\% of symbol period)}
         \label{fig:tone_1}
     \end{subfigure}
     \hspace{0.8cm}
     \begin{subfigure}{0.2\textwidth}
         \centering
         \includegraphics[trim = {65, 185, 60, 205}, clip, height = 3.6cm]{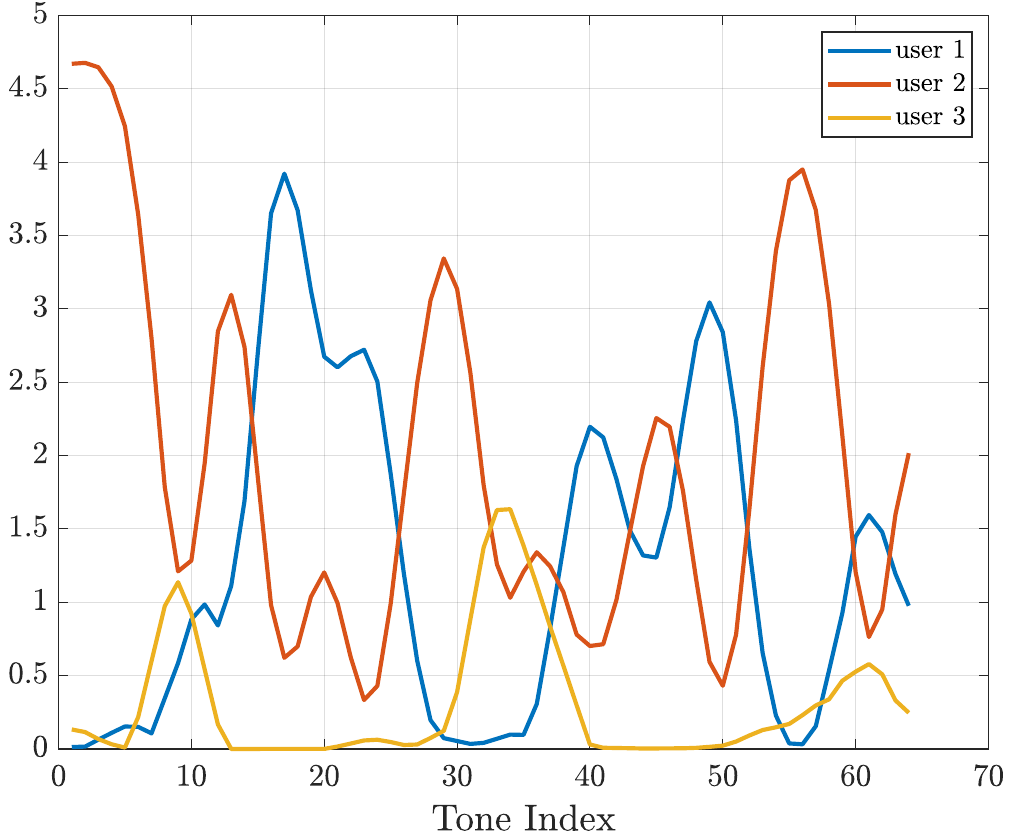}
         \caption{$2^{nd}$ time slot (9\% of symbol period)}
         \label{fig:tone_2}
     \end{subfigure}
     \label{fig:tone}
     \caption{Spectral efficiency versus Tone index for different slots in time domain (SNR = 30dB)}
\end{figure}

Fig.~\ref{fig:tone_1} and~\ref{fig:tone_2} illustrate the effect of the proposed time-sharing algorithm, which leads to users experiencing varying data rates across different time blocks. Specifically, the spectral efficiency, and hence the data rates, for Users 2 and 3 differ between the $1^{\textrm{st}}$ time sharing block and $2^{\textrm{nd}}$ block, while the bit distribution for User 1 remains consistent. 
The first time sharing block comprises the initial 91\% of the symbol period, during which the users' data rates are 123, 170, and 62 Mbps, respectively. On the other hand, for the second time sharing block, which accounts for 9\% of the symbol period, the data rates are 123, 196, and 31 Mbps, respectively. Consequently, the total data rate for a symbol time interval is a weighted summation of their data rates in different time slots, e.g., the average data rate for the second user is $170 \times 0.91 + 196 \times 0.09 = 172.34$ Mbps, while this value for linear receiver is $171.89$ Mbps, which closely aligns with the time-sharing result, with any difference attributable to numerical factors. This demonstrates the proposed solution’s superiority in energy efficiency and data rate performance, particularly in low-rank channel conditions.

\subsection{The effect of increasing number of subcarriers $N$}
\begin{figure}
     \centering
     \hspace{-0.8cm}
     \begin{subfigure}{0.2\textwidth}
         \centering
         \includegraphics[trim = {40, 180, 50, 200}, clip, height = 3.5cm]{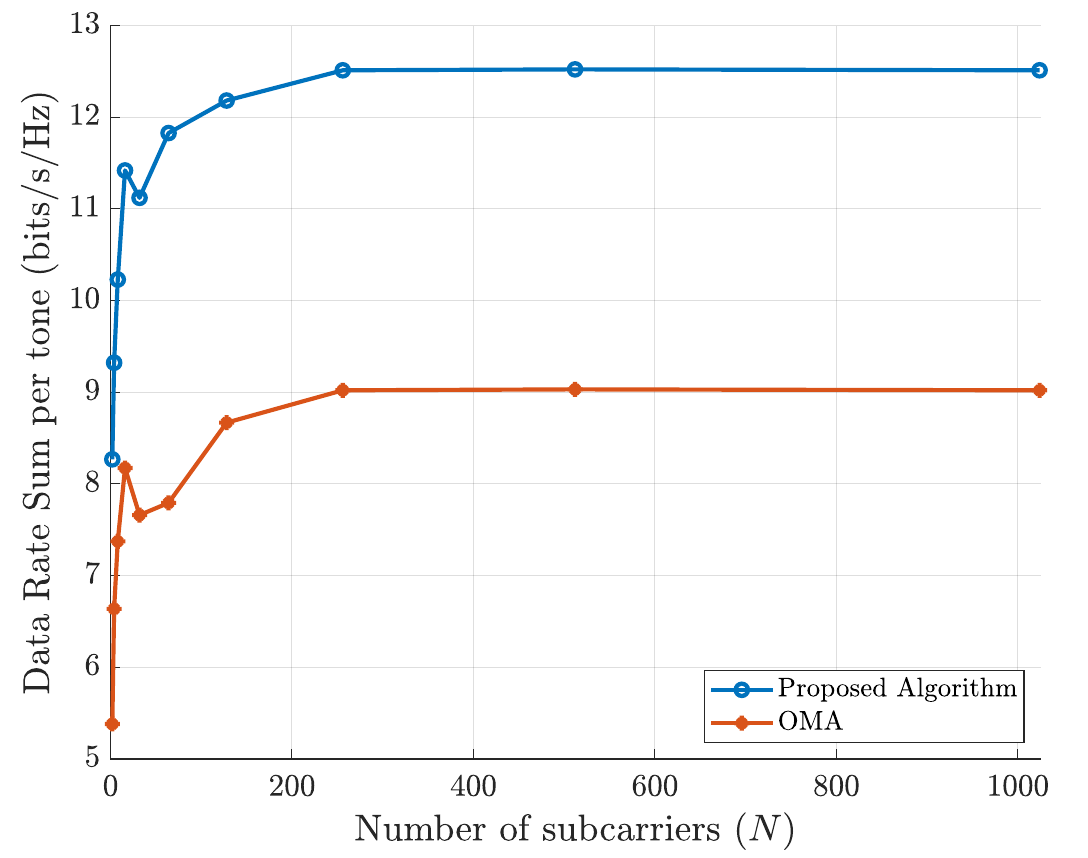}
         \caption{Data rates of three users with distance from BS = $\{500m, 500m, 500m\}$}
         \label{fig:333}
     \end{subfigure}
     \hspace{0.8cm}
     \begin{subfigure}{0.2\textwidth}
         \centering
         \includegraphics[trim = {40, 180, 50, 200}, clip, height = 3.5cm]{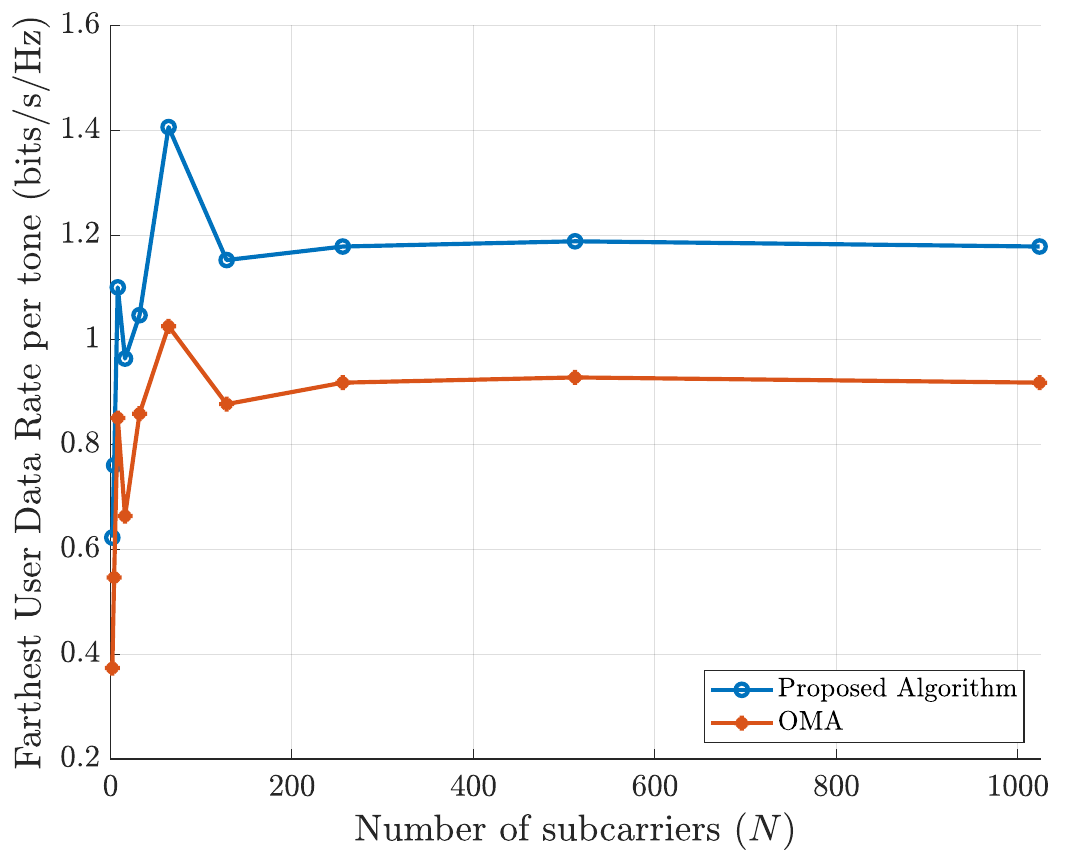}
         \caption{data rate of farthest user with distance from BS = $\{500m, 600m, 700m\}$}
         \label{fig:358}
     \end{subfigure}
     \caption{Data rate sum versus number of subcarriers, single antenna per user, and two antennas at BS}
\end{figure}

Fig. \ref{fig:333} and Fig. \ref{fig:358} demonstrate the effect of increasing the number of subcarriers $N$, on the spectral efficiency, when users are at equal distances of 500m from BS and when their distances from BS are 500m, 600m, 700m, respectively. In this experiment, we compare our proposed algorithm with OMA baseline by setting the receive SNR  as 30dB. Fig. \ref{fig:333} shows the sum of spectral efficiencies across all 3 users as a function of number of subcarriers for both OMA and our proposed algorithm. As we can see, the proposed algorithm outperforms OMA across all numbers of subcarriers. In the case of $N = 1024$, the proposed algorithm achieves $12.51$ bits/s/Hz, while OMA achieves $9.02$ bits/s/Hz.  Fig. \ref{fig:358} shows the spectral efficiency of the farthest user that has the worst channel gain. It is evident that the crosstalk among users decreases as they are going seperate from one another. Even in this case, we see that our proposed algorithm outperform OMA across all numbers of subcarriers.

\section{Conclusion and Future Work}
The proposed power allocation algorithm significantly enhances data rates and energy efficiency  in low-rank channel scenarios. Performance evaluation reveals that this approach, combined with the proposed time sharing algorithm, significantly surpasses current baselines, such as OMA and NOMA, making it an effective solution for multi-user FWA channels. Future work will explore integrating adaptive mechanisms and machine learning to optimize power allocation among users in real time as channel conditions changes. Moreover, since UE and BS are both fixed in FWA systems, new scheduling schemes can be devised for this emerging network architecture.

\footnotesize
\bibliographystyle{IEEEtran}
\bibliography{main.bib}

\end{document}